\documentclass[journal=jctcce,manuscript=article]{achemso}

\usepackage{amsmath}
\usepackage{graphicx}
\usepackage{bm}
\usepackage{lscape}
\usepackage{xcolor}

\SectionNumbersOn

\author{Aleksander L. Wysocki}
\affiliation{Department of Physics, Virginia Tech, Blacksburg, Virginia 24061, United States}
\alsoaffiliation{Present address: Department of Physics and Astronomy, University of Nebraska at Kearney, Kearney, Nebraska 68849}
\email{alexwysocki2@gmail.com}

\author{Kyungwha Park}
\affiliation{Department of Physics, Virginia Tech, Blacksburg, Virginia 24061, United States}
\email{kyungwha@vt.edu}

\title{Relativistic Douglas-Kroll-Hess Calculations of Hyperfine Interactions within First Principles Multireference Methods}

\begin{document}


\begin{abstract}
Relativistic magnetic hyperfine interaction Hamiltonian based on the Douglas-Kroll-Hess (DKH) theory up to the second order is implemented within the \emph{ab initio} multireference methods including spin-orbit coupling in the Molcas/OpenMolcas package.
This implementation is applied to calculate relativistic hyperfine coupling (HFC) parameters for atomic systems and diatomic radicals with valence $s$ or $d$ orbitals by systematically varying active space size in the restricted active space self-consistent field (RASSCF) formalism with restricted active space state interaction (RASSI) for spin-orbit coupling. The DKH relativistic treatment of the hyperfine interaction reduces the Fermi contact contribution to the HFC due to the presence of kinetic factors that regularize the singularity of the Dirac delta function in the nonrelativitic Fermi contact operator. This effect is more prominent for heavier nuclei. As the active space size increases, the relativistic correction of the Fermi contact contribution converges well to the experimental data for light and moderately heavy nuclei. The relativistic correction, however, does not significantly affect the spin-dipole contribution to the hyperfine interaction. In addition to the atomic and molecular systems, the implementation is applied to calculate the relativistic HFC parameters for large trivalent and divalent Tb-based single-molecule magnets (SMMs) such as Tb(III)Pc$_2$ and Tb(II)(Cp$^{\rm{iPr5}})_2$ without ligand truncation using well-converged basis sets. In particular, for the divalent SMM which has an unpaired valence $6s/5d$ hybrid orbital, the relativistic treatment of HFC is crucial for a proper description of the Fermi contact contribution. Even with the relativistic hyperfine Hamiltonian, the divalent SMM is shown to exhibit strong tunability of HFC via an external electric field 
(i.e., strong hyperfine Stark effect).

\end{abstract}

\section{\label{sec:intro}Introduction}


Magnetic hyperfine interaction is an interaction between a nuclear spin moment and an electron magnetic moment. Electronic and magnetic properties of molecules and solids are often investigated by electron paramagnetic resonance\cite{Abragam1970,Atherton1993} and nuclear magnetic resonance\cite{Bertini2001} experiments. An understanding of these resonance spectra requires knowledge of hyperfine interaction and hyperfine coupling (HFC) parameters.
Quantum entanglement, quantum coherence, and quantum control as well as formation of quantum registers based on nuclear spin quantum bits (qubits)\cite{Jelezko2004,Vandersypen2005,vanderSar2012,Wolfowicz2013,Thiele2014,Atzori2018,Godfrin2017,Bourassa2020,Asaad2020,Kundu2022,Ruskuc2022} have been experimentally demonstrated  by manipulating hyperfine interactions within molecules and point defects in solids. For electron-spin qubit systems, however, hyperfine interaction turns out to be one of major sources of spin decoherence.\cite{Takahashi2011,vanderSar2012,Yang2017,Graham2017,Onizhuk2021} Therefore, it is of great importance to understand hyperfine interaction and HFC parameters in various systems.

Calculations of HFC parameters face many challenges. A reliable evaluation of the Fermi contact contribution\cite{Fermi1930} to hyperfine interaction requires a very accurate description of the electronic structure close to and at the nucleus. This involves large, often uncontracted, basis sets, inclusion of the finite-size nucleus, and a proper description of the scalar relativistic effects. In particular, the scalar relativistic effects need to be directly included in the hyperfine Hamiltonian.\cite{Bolvin2017} An accurate treatment of electron correlation must be accompanied for computations of reliable HFC parameters. Indeed, core polarization and higher-order dynamic correlation play a key role in the Fermi contract mechanism.\cite{Tterlikkis1968} For multiconfigurational systems such as lanthanide- or actinide-based magnetic molecules, static correlation must be additionally taken into account especially for a proper evaluation of the paramagnetic spin orbital (PSO) contribution to hyperfine interaction\cite{Wysocki2020,Wysocki2020b,Sharkas2015}.

The majority of first-principles studies of HFC parameters are based on density functional theory (DFT). In this case, the relativistic effects are included both in the electronic structure and the hyperfine interaction Hamiltonian using various approaches including the zeroth-order regular approximation (ZORA),\cite{vanLenthe1998,Belanzoni2001,Autschbach2011,Franzke2022} Douglas-Kroll-Hess (DKH) theory,\cite{Malkin2004} exact two-component method (X2C),\cite{Autschbach2017,Franzke2022} or the four-component Dirac Hamiltonian.\cite{Malkin2011} Finite-nucleus models can be also used for the electrostatic\cite{Aquino2012} and/or magnetic nuclear potentials.\cite{Malkin2006} DFT works primarily for single-reference systems since
it provides a decent treatment of dynamic correlation. An alternative single-reference approach is the orbital-optimized second-order Møller–Plesset perturbation theory (OOMP2).\cite{Sandhoefer2013} Much more accurate treatment of dynamic correlation in single-reference systems is, however, provided by the coupled cluster (CC) technique. Indeed, CC calculations of HFC based on the Dirac Hamiltonian provide theoretical benchmarks.\cite{Martensson2002,Sahoo2003,Sasmal2015,Haase2020} Such calculations are, however, quite expensive and are feasible only for atoms and relatively small molecules.

Multireference quantum chemistry techniques are ideal methods to reliably calculate HFC parameters for multiconfigurational systems where static correlation needs to be addressed. This is especially true for systems with orbitally (quasi-)degenerate ground states for which a proper account of static correlation is crucial for a description of the orbital angular momentum and the PSO contribution to HFC. There is a number of reported multireference studies of HFC parameters.\cite{Fernandez1992,Jonsson2007,Song2007,Bieron2009,Lan2014,Sharkas2015,Lan2015,Shiozaki2016,Samanta2018,Wysocki2020,Feng2021,Birnoschi2022} This includes multiconfigurational methods,\cite{Jonsson2007,Song2007,Bieron2009} complete active space self consistent field (CASSCF) and restricted active space self consistent field (RASSCF) calculations,\cite{Fernandez1992,Lan2014,Sharkas2015,Lan2015,Wysocki2020,Feng2021,Birnoschi2022} CASPT2 method,\cite{Shiozaki2016} and multirefrence CC theory.\cite{Samanta2018} Although many of these studies have employed nonrelativistic hyperfine interaction Hamiltonian,\cite{Fernandez1992,Lan2014,Sharkas2015,Shiozaki2016,Samanta2018,Wysocki2020} there are several works that have used relativistic hyperfine interaction Hamiltonian based on the Dirac Hamiltonian,\cite{Jonsson2007,Song2007,Bieron2009} the DKH formalism,\cite{Lan2015} or the X2C treatment.\cite{Feng2021,Birnoschi2022} 

Here we present our implementation of the relativistic DKH hyperfine interaction Hamiltonian within the Molcas/OpenMolcas package\cite{Molcas,Openmolcas} and use it to perform CASSCF/RASSCF calculations of HFC parameters with restricted active space state interaction (RASSI) for spin-orbit coupling for various compounds including atomic systems, diatomic radicals, and large Tb-based single-molecule magnets (SMMs). We analyze the effects of dynamic correlation, finite-nucleus models for nuclear electrostatic potentials, and the relativistic treatment of hyperfine interaction Hamiltonian for the considered compounds. Our results are also compared to the experimental data and results from other computational methods.

\section{\label{sec:theory}Theory}

In this section we present the theory of magnetic hyperfine interaction within the DKH treatment of relativistic effects. A similar formalism is discussed in Refs.~\citenum{Dyall2000,Fukuda2003,Melo2005,Sandhoefer2013,Lan2015}. In the following we use Hartree atomic units. Quantum-mechanical operators are denoted by hats. Vectors, tensors and matrices are written in a bold font.

We consider a point nucleus located at the origin with atomic number $Z$ and magnetic moment $\hat{\mathbf{m}}_N=g_N\mu_N\hat{\mathbf{I}}$, where $g_N$ is the nuclear $g$-factor, $\mu_N=\frac{1}{2M_p}$ is the nuclear magneton ($M_p$ is the proton mass), and $\hat{\mathbf{I}}$ is the nuclear spin. The nuclear spin produces a magnetic field associated with the following magnetic vector potential
\begin{equation}
\hat{\mathbf{A}}_N=\alpha^2\frac{\hat{\mathbf{m}}_N\times\mathbf{r}}{r^3},
\end{equation}
where $\alpha$ is the fine structure constant and $\mathbf{r}$ is the radial vector from the nucleus. This nuclear magnetic field interacts with the electrons and the microscopic interaction Hamitonian is denoted by $\hat{H}_{\text{MHf}}$.

Let us consider an electronic multiplet composed of $N$ electronic states that can be described as a fictitious pseudo-spin $S_\text{eff}$. When the energy separation of the multiplet from the rest of electronic states is much larger than the energy scale of $\hat{H}_{\text{MHf}}$ ($\sim$ 0.1 cm$^{-1}$), the magnetic hyperfine interaction within the subspace of this multiplet can be mapped into pseudo-spin Hamiltonian such as 
\begin{equation}
\hat{H}_{A}=\hat{\mathbf{I}}\cdot\mathbf{A}\cdot\hat{\mathbf{S}}_\text{eff}.
\end{equation}
Here $\mathbf{A}$ is a $3\times3$ magnetic hyperfine matrix that can be found using the following formula\cite{Abragam1970,Chibotaru2012,Feng2021}
\begin{equation}
(\mathbf{A}\mathbf{A}^T)_{\alpha\beta}=\frac{3}{S_\text{eff}(S_\text{eff}+1)(2S_\text{eff}+1)}\sum_{ij}^N\langle i|\hat{h}_{\text{MHf}}^\alpha|j\rangle\langle j|\hat{h}_{\text{MHf}}^\beta|i\rangle,
\label{eq:AAT-tensor}
\end{equation}         
where $\alpha, \beta=x,y,z$ and $\hat{h}_{\text{MHf}}^\alpha\equiv\partial\hat{H}_{\text{MHf}}/\partial\hat{I}_\alpha$. Here the summation runs over the states of the electronic multiplet. By appropriate rotation, the $\mathbf{A}\mathbf{A}^T$ tensor can be transformed into a diagonal form. Taking the square root of its diagonal values, we obtain eigenvalues of the $\mathbf{A}$ matrix. The full $\mathbf{A}$ matrix can be then obtained from its diagonal form by the inverse rotation. Note that this procedure determines the eigenvalues of the $\mathbf{A}$ matrix up to a sign. The sign of the hyperfine matrix elements can be inferred from other physical and chemical considerations or experiments. 

As shown in Eq.(~\ref{eq:AAT-tensor}), the calculation of the $\mathbf{A}$ matrix requires knowledge of the microscopic magnetic hyperfine interaction Hamiltonian $\hat{H}_{\text{MHf}}$. Its specific form depends on the treatment of relativistic effects. We discuss the relativistic treatment and hyperfine Hamiltonian that we implement below.

\subsection{\label{sec:nrel}Nonrelativistic Theory}

In the nonrelativistic treatment,\cite{Ramsey1950} $\hat{H}_{\text{MHf}}$ can be derived by considering an electron with spin magnetic moment $\hat{\mathbf{m}}_s=-g_s\mu_B\hat{\boldsymbol{\sigma}}/2$ ($g_s\approx2$ is the electronic g-factor, $\mu_B=\frac{1}{2}$ is the Bohr magneton, and $\hat{\sigma}$ are the Pauli matrices) moving with momentum $\hat{\mathbf{p}}=-i\nabla$ in the presence of a nuclear magnetic field. The nonrelativistic Hamiltonian for such a system is given by

\begin{equation}
\hat{H}=\frac{1}{2}\left(\hat{\mathbf{p}}+\hat{\mathbf{A}}_\text{N}\right)^2 -\hat{\mathbf{m}}_s\cdot\{\nabla\times\hat{\mathbf{A}}_\text{N}\},
\end{equation}
where the derivative operator acts only inside the curly bracket. The magnetic hyperfine interaction Hamiltonian corresponds to the linear terms in $\hat{\mathbf{A}}_\text{N}$ in the above equation. It can be re-written as a sum of three contributions
\begin{equation}
\hat{H}^{\text{NREL}}_{\text{MHf}}=\hat{H}^{\text{NREL}}_{\text{PSO}}+\hat{H}^{\text{NREL}}_{\text{SD}}+\hat{H}^{\text{NREL}}_{\text{FC}},
\label{NR:HMHf}
\end{equation}
where the first term represents a magnetostatic interaction between $\hat{\mathbf{m}}_N$ and the electron orbital magnetic moment. This so called paramagnetic spin-orbital (PSO) contribution is given by
\begin{equation}
\hat{H}^{\text{NREL}}_\text{PSO}=\alpha^2\frac{\hat{\mathbf{l}}\cdot\hat{\mathbf{m}}_N}{r^3},
\label{NR:HPSO}
\end{equation}
where $\hat{\mathbf{l}}$ is the electronic orbital angular momentum operator. The second term in Eq.~(\ref{NR:HMHf}) represents the anisotropic through-space dipolar interaction between $\hat{\mathbf{m}}_N$ and $\hat{\mathbf{m}}_s$. This spin-dipole contribution is given by
\begin{equation}
\hat{H}^{\text{NREL}}_{\text{SD}}=\alpha^2\frac{1}{r^3}\bigg[\hat{\mathbf{m}}_s\cdot\hat{\mathbf{m}}_N - 3\frac{(\hat{\mathbf{m}}_s\cdot\mathbf{r})(\hat{\mathbf{m}}_N\cdot\mathbf{r})}{r^2}\bigg].
\end{equation}
Finally, the last term in Eq.~(\ref{NR:HMHf}) represents the contact interaction between $\hat{\mathbf{m}}_N$ and the electronic spin density at the nucleus position.\cite{Fermi1930} This isotropic Fermi contact term is given by
\begin{equation}
\hat{H}^{\text{NREL}}_{\text{FC}}=-\frac{8\pi}{3}\alpha^2\hat{\mathbf{m}}_s\cdot\hat{\mathbf{m}}_N\delta(\mathbf{r}).
\end{equation}
The presence of a Dirac delta function $\delta(\mathbf{r})$ in the above expression is an artifact of the nonrelativistic treatment of the hyperfine interaction (and the point-nucleus model for the nuclear magnetic vector potential). Since a (quasi-)relativistic electronic wavefunction shows a logarithmic divergence at the center of a nucleus, matrix elements of the nonrelativistic Fermi contact contribution become singular. Although the use of an incomplete basis set or a finite-nucleus model may lead to finite matrix elements, these values would exhibit an unphysical dependence on the size of the basis set and the nuclear model. In particular, such an approach would manifest lack of basis-set convergence of the Fermi contact interaction parameter. Therefore, utilization of the nonrelativistic hyperfine Hamiltonian combined with a (quasi-)relativistic electronic wavefunction cannot reliably evaluate the Fermi contact term.

\subsection{\label{sec:dkh}DKH}

In order to include the relativistic effects in the magnetic hyperfine interaction, we use the DKH theory.\cite{Douglass1974,Hess1986,Wolf2002,Reiher2004} We consider the following four-component Hamiltonian
\begin{equation}
\hat{H}_{4c}=c^2 \hat{\boldsymbol{\beta}} +c\hat{\boldsymbol{\alpha}}\cdot\hat{\mathbf{p}}+V_N+c\hat{\boldsymbol{\alpha}}\cdot\hat{\mathbf{A}}_N,
\label{H4c}
\end{equation}
where $c$ and $V_N$ are the speed of light and the nuclear electrostatic potential, respectively, and 
\begin{equation}
 \hat{\boldsymbol{\alpha}} =
 \begin{pmatrix}
  0 & \hat{\boldsymbol{\sigma}} \\
  \hat{\boldsymbol{\sigma}} & 0
  \end{pmatrix}, \qquad
 \boldsymbol \hat{\boldsymbol{\beta}}=
 \begin{pmatrix}
  \boldsymbol I_2 & 0 \\
  0 & -\boldsymbol I_2
  \end{pmatrix}. 
\end{equation}
Here $\boldsymbol I_2$ is a $2\times2$ identity matrix. In Eq.~(\ref{H4c}) the first three terms constitute the Dirac Hamiltonian, and the last term describes the magnetic hyperfine interaction.
In the DKH method we apply a series of unitary transformation to the four-component Hamiltonian such that the off-diagonal blocks become successively smaller and can be eventually neglected. Then the upper $2\times2$ block can be taken as an effective two-component Hamiltonian.

Originally, this procedure was carried out for the Dirac Hamiltonian.\cite{Douglass1974,Hess1986} The successive unitary transformations on the Dirac Hamiltonian produce off-diagonal terms in increasing powers of a damped external potential (which is defined to be $V_N$ divided by a factor of the order of $c^2$). Therefore, this procedure converges rather rapidly. An effective $2\times2$ DKH Hamiltonian numerically equivalent to the original $4\times4$ Dirac Hamiltonian, is then obtained for a description of electronic states. In practice, it is usually sufficient to perform only two unitary transformations.\cite{Wolf2002,Reiher2004} In the presence of a nuclear magnetic field, there are several ways to proceed. Dyall generalized the DKH formalism by replacing the dynamical momentum $\hat{\mathbf{p}}$ by the canonical momentum $\hat{\mathbf{p}}+\hat{\mathbf{A}}_N$ in the unitary transformations.\cite{Dyall2000} Malkin \emph{et al.}, employed the DKH formalism for the Dirac Hamiltonian and evaluated an expectation value of the four-component hyperfine Hamiltonian (the last term in Eq.~(\ref{H4c})) using the four-component spinor wavefunction obtained from the DKH two-component wavefunction by applying DKH unitary transformations backwards.\cite{Malkin2004} Finally, Fukuda \emph{et al.}, applied DKH unitary transformations to Eq.~(\ref{H4c}) by treating the nuclear magnetic field on equal footing with electrostatic electron-nucleus interaction.\cite{Fukuda2003} Here, we adapt the approach by Fukuda \emph{et al.}(Ref.~\citenum{Fukuda2003}).

\subsubsection{\label{sec:dkh1}DKH1}

The first (or zeroth-order) unitary transformation is the free-particle Foldy–Wouthuysen (FW) transformation described by the following $4\times4$ unitary matrix\cite{Foldy1950}
\begin{equation}
\hat{U}_0=\hat{K}\Big(1+\hat{\beta} \hat{R}\hat{\boldsymbol{\alpha}}\cdot\hat{\mathbf{p}}\Big),
\end{equation}
where
\begin{equation}
\hat{K}=\sqrt{\frac{\hat{\mathcal{E}}_p+c^2}{2\hat{\mathcal{E}}_p}},
\end{equation}
\begin{equation}
\hat{R}=\frac{c}{\hat{\mathcal{E}}_p+c^2},
\end{equation}
and
\begin{equation}
\hat{\mathcal{E}}_p=c\sqrt{\hat{p}^2+c^2}.
\end{equation}
Applying the FW transformation to the Hamiltonian Eq.~(\ref{H4c}) we obtain
\begin{equation}
\hat{H}^{(1)}_{4c}=\hat{U}_0\hat{H}_{4c}\hat{U}_0^\dagger=\hat{E}_0+\hat{E}_1+\hat{O}_1,
\end{equation}
where
\begin{equation}
\hat{E}_0=\hat{\beta}\hat{\mathcal{E}}_p,
\end{equation}
\begin{equation}
\hat{E}_1=\hat{E}^V_1+\hat{E}^A_1,
\end{equation}
\begin{equation}
\hat{O}_1=\hat{O}_1^V+\hat{O}_1^A.
\end{equation}
Here $E$ symbols denote even $2\times2$ block diagonal terms and $O$ symbols denote odd $2\times2$ block off-diagonal terms. Subscripts in the above expressions denote the order of $V_N$. The terms denoted by superscript $V$ originate from the FW transformation applied to $V_N$ and are given by

\begin{equation}
\hat{E}_1^V=\hat{K}\Big[V_N+\hat{R}\big(\hat{\boldsymbol{\alpha}}\cdot\hat{\mathbf{p}}\big)V_N\big(\hat{\boldsymbol{\alpha}}\cdot\hat{\mathbf{p}}\big)\hat{R}\Big]\hat{K},
\end{equation}
\begin{equation}
\hat{O}_1^V=\hat{\beta}\hat{K}\Big[\hat{R}\big(\hat{\boldsymbol{\alpha}}\cdot\hat{\mathbf{p}}\big)V_N-V_N\big(\hat{\boldsymbol{\alpha}}\cdot\hat{\mathbf{p}}\big)\hat{R}\Big]\hat{K}.
\end{equation}
The terms denoted by superscript $A$ originate from the FW transformation applied to the last term in Eq.~(\ref{H4c}) and are given by
\begin{equation}
\hat{E}^{A}_1=c\hat{\beta} \hat{K}\Big[\hat{R}\big(\hat{\boldsymbol{\alpha}}\cdot\hat{\mathbf{p}}\big)\big(\hat{\boldsymbol{\alpha}}\cdot\hat{\mathbf{A}}_N\big)+\big(\hat{\boldsymbol{\alpha}}\cdot\hat{\mathbf{A}}_N\big)\big(\hat{\boldsymbol{\alpha}}\cdot\hat{\mathbf{p}}\big)\hat{R}\Big]\hat{K}
\end{equation}
\begin{equation}
\hat{O}^{A}_1= c\hat{K}\Big[\big(\hat{\boldsymbol{\alpha}}\cdot\hat{\mathbf{A}}_N\big)-\hat{R}\big(\hat{\boldsymbol{\alpha}}\cdot\hat{\mathbf{p}}\big)\big(\hat{\boldsymbol{\alpha}}\cdot\hat{\mathbf{A}}_N\big)\big(\hat{\boldsymbol{\alpha}}\cdot\hat{\mathbf{p}}\big)\hat{R}\Big]\hat{K}.
\end{equation}

In the DKH1 approximation, the effective two-component Hamiltonian is given by the upper $2\times2$ diagonal block of $\hat{H}^{(1)}_{4c}$. Note that this results in neglecting the odd $\hat{O}_1$ term. Consequently, in the DKH1 approximation an effective $2\times2$ magnetic hyperfine Hamiltonian is given by an upper $2\times2$ diagonal block of $\hat{E}^{A}_1$. Using the following identity
\begin{equation}
\big(\boldsymbol{\alpha}\cdot\hat{\mathbf{a}}\big)\big(\boldsymbol{\alpha}\cdot\hat{\mathbf{b}}\big)=\hat{\mathbf{a}}\cdot\hat{\mathbf{b}}+i\boldsymbol{\sigma}\cdot\big(\hat{\mathbf{a}}\times\hat{\mathbf{b}}\big),
\label{identity}
\end{equation}
for any two operators $\hat{\mathbf{a}}$ and $\hat{\mathbf{b}}$, we obtain 
\begin{equation}
\hat{H}^{\text{DKH1}}_{MHf}=\hat{H}^{\text{DKH1}}_{PSO}+\hat{H}^{\text{DKH1}}_{FC+SD},
\end{equation}
where the spin-independent PSO term is given by
\begin{equation}
\hat{H}^{\text{DKH1}}_{PSO}=c\hat{K}\Big[\hat{R}\big(\hat{\mathbf{p}}\cdot\hat{\mathbf{A}}_N\big)+\big(\hat{\mathbf{A}}_N\cdot\hat{\mathbf{p}}\big)\hat{R}\Big]\hat{K},
\label{hmhf-pso-dkh1}
\end{equation}
while the spin-dependent term is given by
\begin{equation}
\hat{H}^{\text{DKH1}}_{FC+SD}=ic\boldsymbol{\sigma}\cdot \hat{K}\Big[\hat{R}\big(\hat{\mathbf{p}}\times\hat{\mathbf{A}}_N\big)+\big(\hat{\mathbf{A}}_N\times\hat{\mathbf{p}}\big)\hat{R}\Big]\hat{K}.
\label{hmhf-fcsd-dkh1}
\end{equation}
$\hat{H}^{\text{DKH1}}_{\text{FC+SD}}$ represents a sum of the Fermi contact and spin-dipole contributions which cannot be naturally separated in the relativistic theory. More importantly, the presence of $\hat{K}$ and $\hat{R}$ kinematic factors regularizes the magnetic hyperfine Hamiltonian in the vicinity of the nucleus so that singularities do not arise as in the nonrelativistc theory. Note that in the nonrelativistic limit ($c\rightarrow\infty$), we obtain $\hat{K}=1$ and $\hat{R}=1/2c$. Then the DKH1 equations recover the nonrelativistic expressions.

\subsubsection{\label{sec:dkh1}DKH2}

In the DKH2 approximation, we take the aforementioned four-component Hamiltonian $\hat{H}^{(1)}_{4c}$ and apply to it a second unitary transformation. Following the generalized DKH theory,\cite{Wolf2002,Reiher2004} the corresponding unitary matrix is given by
\begin{equation}
\hat{U}_1=1+\hat{W}_1+\frac{1}{2}\hat{W}_1^2,
\end{equation}
where $\hat{W}_1$ is an odd anti-Hermitian operator ($\hat{W}_1^\dagger=-\hat{W}_1$) that is first order in $V_N$ and $\hat{\mathbf{A}}_N$ potentials. In the DKH2 approximation, the terms of third and higher orders in these potentials are neglected and the transformed Hamiltonian is given by
\begin{equation}
\hat{H}^{(2)}_{4c}=\hat{U}_1\hat{H}^{(1)}_{4c}\hat{U}_1^\dagger=\hat{E}_0+\hat{E}_1+\hat{O}_1+[\hat{W}_1,\hat{E}_0]+[\hat{W}_1,\hat{O}_1]
+\frac{1}{2}[\hat{W}_1,[\hat{W}_1,\hat{E}_0]]+[\hat{W}_1,\hat{E}_1],
\end{equation}
where $[,]$ denotes a commutator. The first two terms correspond to the DKH1 Hamiltonian, and the third and fourth terms are odd first-order terms in the potentials. The fifth and sixth terms are even 
second-order terms in the potentials, and the last term is an odd second-order term in the potentials. The operator $\hat{W}_1$ is chosen in such a way that the odd first-order term in the potentials vanishes.
\begin{equation}
\hat{O}_1+[\hat{W}_1,\hat{E}_0]=0.
\label{DKH2condition}
\end{equation}
It is convenient to decompose $\hat{W}_1$ into part proportional to $V_N$ and part proportional to $\hat{\mathbf{A}}_N$
\begin{equation}
\hat{W}_1=\hat{W}_1^V+\hat{W}_1^A.
\end{equation}
The $\hat{W}_1^{A,V}$ operators can be evaluated in the $p$-basis (space of eigenfunctions of the magnitude of the momentum operator) where $\hat{\mathcal{E}}_p$ is diagonal
\begin{equation}
(\hat{W}^V_1)_{pp'}=\hat{\beta}\frac{(\hat{O}^V_1)_{pp'}}{\hat{\mathcal{E}}_p+\hat{\mathcal{E}}_{p'}},
\end{equation}
\begin{equation}
(\hat{W}^A_1)_{pp'}=\hat{\beta}\frac{(\hat{O}^A_1)_{pp'}}{\hat{\mathcal{E}}_p+\hat{\mathcal{E}}_{p'}},
\end{equation}
where we use the fact that $\hat{W}_1^{A,V}$ operators (being odd in $\hat{\boldsymbol{\alpha}}$) anticommute with $\hat{\beta}$.

Using Eq. (\ref{DKH2condition}) we obtain the following expression for the DKH2-transformed four-component Hamiltonian
\begin{equation}
\hat{H}^{(2)}_{4c}=\hat{E}_0+\hat{E}_1+\hat{E}_2+\hat{O}_2,
\end{equation}
where
\begin{equation}
\hat{E}_2=\frac{1}{2}[\hat{W}_1,\hat{O}_1],
\end{equation}
\begin{equation}
\hat{O}_2=[\hat{W}_1,\hat{E}_1].
\end{equation}
In the DKH2 approximation, the $\hat{O}_2$ term is neglected and the effective two-component Hamiltonian is given by an upper $2\times2$ block of $\hat{H}^{(2)}_{4c}$. The $\hat{\mathbf{A}}_{\rm{N}}$-dependent part of this effective Hamiltonian is given by
\begin{equation}
\hat{E}_1^A+\frac{1}{2}[W^V,\hat{O}_1^A]+\frac{1}{2}[W^A,\hat{O}_1^V]+\frac{1}{2}[W^A,\hat{O}_1^A].
\label{dkh2-Adep}
\end{equation}
Note that the last term in the above expression is of second order in the magnetic vector potential and so it describes the diamagnetic effect. Although the diamagnetic term may, in principle, contribute to the hyperfine interaction (see, for example, Ref~\citenum{Sandhoefer2013}), we neglect such a contribution in this work.

The magnetic hyperfine Hamiltonian in the DKH2 approximation is given by an upper $2\times2$ diagonal block of Eq.~(\ref{dkh2-Adep})
\begin{equation}
\hat{H}^{\text{DKH2}}_{MHf}=\hat{H}^{\text{DKH1}}_{MHf}+\frac{1}{2}[o_V,w_A]+\frac{1}{2}[w_V,o_A],
\label{Hdkh2-MHf}
\end{equation}
where
\begin{equation}
\hat{o}_V=\hat{K}\Big[\hat{R}\big(\hat{\boldsymbol{\sigma}}\cdot\hat{\mathbf{p}}\big)V_N-V_N\big(\hat{\boldsymbol{\sigma}}\cdot\hat{\mathbf{p}}\big)\hat{R}\Big]\hat{K},
\end{equation}
\begin{equation}
\hat{o}_A= c\hat{K}\Big[\big(\hat{\boldsymbol{\sigma}}\cdot\hat{\mathbf{A}}_N\big)-\hat{R}\big(\hat{\boldsymbol{\sigma}}\cdot\hat{\mathbf{p}}\big)\big(\hat{\boldsymbol{\sigma}}\cdot\hat{\mathbf{A}}_N\big)\big(\hat{\boldsymbol{\sigma}}\cdot\hat{\mathbf{p}}\big)\hat{R}\Big]\hat{K},
\end{equation}
\begin{equation}
(\hat{w}_V)_{pp'}=\frac{(\hat{o}_V)_{pp'}}{\hat{\mathcal{E}}_p+\hat{\mathcal{E}}_{p'}},
\end{equation}
\begin{equation}
(\hat{w}_A)_{pp'}=\frac{(\hat{o}_A)_{pp'}}{\hat{\mathcal{E}}_p+\hat{\mathcal{E}}_{p'}}.
\label{wapp}
\end{equation}
Using the identity Eq.~(\ref{identity}), the magnetic hyperfine Hamiltonian can be separated into the spin-independent (PSO) and spin-dependent (FC+SD) parts
\begin{equation}
\hat{H}^{\text{DKH2}}_{MHf}=\hat{H}^{\text{DKH1}}_{PSO}+\hat{H}^{\text{DKH2}}_{PSO}+\hat{H}^{\text{DKH1}}_{FC+SD}+\hat{H}^{\text{DKH2}}_{FC+SD}.
\end{equation}
The expressions for $\hat{H}^{\text{DKH2}}_{PSO}$ and $\hat{H}^{\text{DKH2}}_{FC+SD}$ are rather lengthy and are shown in the Supporting Information (SI).

\section{\label{sec:comp}Implementation and Computational Details}

We implement calculations of the hyperfine interaction matrix, $\mathbf{A}$, within the Molcas/OpenMolcas package\cite{Molcas,Openmolcas} by using the aforementioned DKH formalism. The key difference with the previous nonrelativistic implementation\cite{Sharkas2015,Wysocki2020} is the computation of matrix elements of $\hat{h}_{\text{MHf}}^\alpha$ based on the $\hat{H}^{\text{DKH1}}_\text{MHf}$ and $\hat{H}^{\text{DKH2}}_\text{MHf}$ Hamiltonians (see Eqs.~\ref{hmhf-pso-dkh1}-\ref{hmhf-fcsd-dkh1} and Eqs.~S1-S4 in the SI). Using the the resolution of identity method developed by Hess \emph{et al.},\cite{Hess1986b} the matrix elements can be expressed in terms of $\hat{p}^\alpha\frac{r^\beta}{r^3}$, $\frac{r^\beta}{r^3}\hat{p}^\alpha$, and $\hat{p}^\alpha V_N\hat{p}^\beta$ integrals. These integrals are evaluated with the Rys quadrature method\cite{Lindh1991} using the turn-over-rule, which allows one to act the derivative ($\hat{p}^\alpha$) operator on the Gaussian basis functions rather than evaluating derivatives of $V_N$ or $\frac{r^\beta}{r^3}$. As in the standard DKH procedure, the integrals are then transformed to the $p$-basis, where all the kinematic factors ($\hat{K}$, $\hat{R}$, $\hat{M}$, $\hat{N}$, $\hat{L}$) are diagonal, and thus they can be easily evaluated. The matrix elements of $\hat{h}_{\text{MHf}}^\alpha$ are then 
computed and back-transformed to the original coordinate basis.

Calculations of HFC parameters require very good basis sets due to a need for an accurate description of the region near the nucleus. As discussed above, the DKH formalism relies on the use of the completeness relation (resolution of identity) in the evaluations of the relativistic hyperfine operator which is valid only for very large basis sets. Therefore, for atomic systems and diatomic radicals, we use fully uncontracted relativistic atomic natural orbital (ANO-RCC) basis sets.\cite{Widmark1990,Roos2004}
For large molecules like TbPc$_2$ and TbCp(Cp$^{\rm{iPr5}})_2$  such basis sets are, however, computationally too demanding. In this case, the electronic structure calculations are performed using contracted ANO-RCC basis sets. In particular, polarized valence triple-$\zeta$ quality (ANO-RCC-VTZP) is used for the Tb atoms, polarized valence double-$\zeta$ quality (ANO-RCC-VDZP) is used for the nitrogen and the carbon atoms, and valence double-$\zeta$ quality (ANO-RCC-VDZ) is used for the hydrogen atoms. However, the DKH calculations of the relativistic Hamiltonian and hyperfine operators are performed using fully uncontracted (primitive) basis sets. For all of the considered systems, scalar relativistic effects in the Born-Oppenheimer Hamiltonian are also included within the DKH2 approximation.\cite{Douglass1974,Hess1986}

Calculations of the electronic structure are performed by following two steps. First, in the absence of spin-orbit coupling, the spin-free energies and eigenstates are obtained for a given spin-multiplicity using either complete active space self-consistent field (CASSCF)~\cite{Roos1980,Siegbahn1981} or restricted active space self-consistent field (RASSCF)\cite{Malmqvist1990} methods. In the second step, the SOC is included within the atomic mean-field approximation~\cite{Hess1996} using the restricted active space state interaction (RASSI)~\cite{Malmqvist2002} method. 

\section{\label{sec:results}Results and Discussion}

\subsection{\label{sec:atoms}Atomic systems}

Atoms and ions are ideal systems to test the accuracy of our relativistic HFC calculations. In this case, the HFC tensor is fully isotropic and can be described by a single parameter $A_\text{iso}=\text{Tr}(\mathbf{A})/3$ that can be extracted from experiments. We focus on HFC for the group II cations and group XI neutral atoms. These are spin-$1/2$ radicals with a valence $s$ orbital in the orbitally nondegenerate ground state so that the hyperfine interaction originates solely from the Fermi contact mechanism that is particularly sensitive to the relativistic effects.

\begin{figure}[t]
\centering
\includegraphics[width=1.0\linewidth]{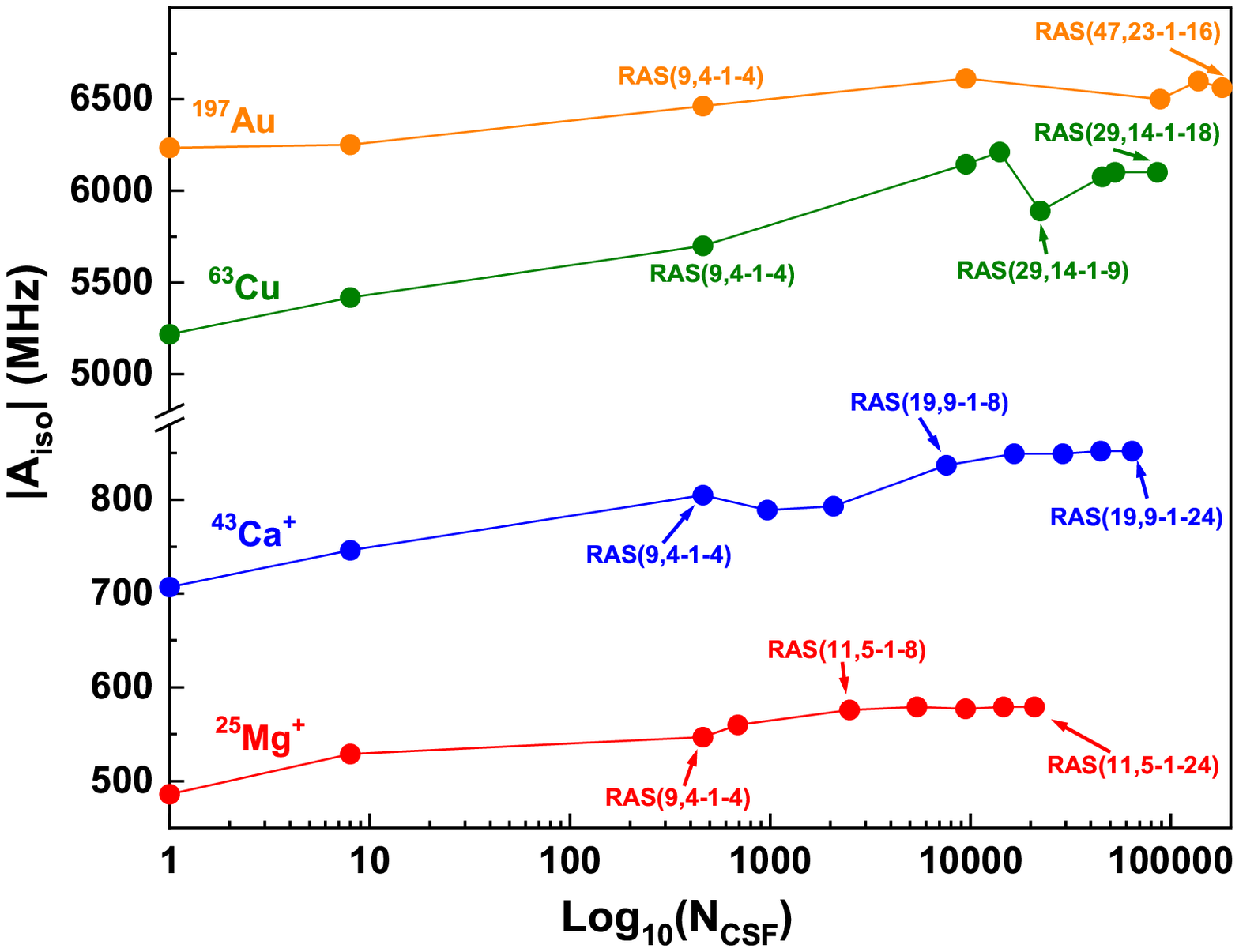}
\caption{Magnitude of the isotropic HFC constant for several of group II cations and group X neutral atoms calculated using RASSCF method as a function of number of configuration state functions. DKH2 relativistic hyperfine Hamiltonian and point nucleus electrostatic nuclear potential are used in calculations. The active spaces corresponding to some of the data points is indicated.}
\label{DynCorrel}
\end{figure}

A proper treatment of dynamical correlation is crucial for accurate calculations of hyperfine interaction \cite{Chipman1992,Neese2009,Sandhoefer2013}. In particular, spin polarization of core-shell $s$ orbitals is known to have a strong effect on the Fermi contact contribution.\cite{Abragam1955,Watson1961} Here, we use the RASSCF method that allows for an efficient treatment of the core spin polarization as well as higher-order correlations. The active space is denoted as RAS($N_e$,$N_1$-$N_2$-$N_3$) where $N_e$ is the number of electrons in the active space, $N_1$ is the number of RAS1 orbitals, $N_2$ is the number of RAS2 orbitals, and $N_3$ is the number of RAS3 orbitals. The RAS2 space consists of only the outermost unpaired-spin electron ($N_2$=1), while $N_1$ and $N_3$ are varied. We allow for two-electron and two-hole excitations in RAS3 and and RAS1 spaces, respectively. For small values of $N_1$ and $N_3$, we confirm that the RASSCF calculations agree well with corresponding complete active space self-consistent field (CASSCF) calculations. Starting from the restricted open-shell Hartree Fock (ROHF) calculations ($N_1=N_3=0$), we systematically expand the active space in three stages. First, the most important correlations are included by successively adding pairs of same symmetry highest-orbital-energy inactive and lowest-orbital-energy secondary shells to RAS1 and RAS3 spaces, respectively. Second, all core-polarization effects are included by successively adding remaining occupied shells to RAS1 space. This stage is completed for all atoms and cations except for Au in which case inclusion of all occupied orbitals in RAS1 space is computationally unfeasible. Therefore, for Au only one of the occupied $d$ shells is included in the RAS1 space. Finally, in the third stage, for not too heavy atoms and ions (Mg$^+$, Ca$^+$, Sr$^+$, and Cu), additional higher-order correlations are included by successively adding empty orbitals to the RAS3 space.

Figure~\ref{DynCorrel} shows the calculated magnitude of $A_\text{iso}$ for several group II cations and group X neutral atoms as a function of number of configuration state functions used in the RASSCF calculations. Here, we use the DKH2 relativistic hyperfine Hamiltonian and the point nucleus electrostatic nuclear potential. In general, inclusion of dynamic correlation increases the $|A_\text{iso}|$ value compared to that from the ROHF. This can be explained by the correlation-induced contraction of core electrons toward the nucleus, which causes an increase of the Fermi contact contribution.\cite{Filatov2004} For Mg$^+$ and Ca$^+$, all nominally doubly occupied orbitals are included in RAS1 and $A_\text{iso}$ appears well converged with respect to the number of empty orbitals in RAS1 (see Tables S1-S2 in the SI). The similar behavior is observed for Sr$^+$ and Ba$^+$ (see Tables S3-S4 in the SI). In the case of Cu, a much stronger dependence on the size of the active space is observed. Nevertheless, for larger active spaces the convergence seems to be reached. For Au and Ag (see Tables S5-S6 in the SI), while $A_\text{iso}$ shows a rather weak dependence on the size of the active space, the full convergence is not achieved and even for the largest active spaces that we consider, noticeable variations are observed.

\begin{table}[t]
\centering
\caption{Magnitude of the calculated isotropic HFC parameter $A_\text{iso}$ (in MHz) for atomic systems compared with experimental data and other calculations in the literature (Lit.). PN and FN stand for the point-nucleus and the finite-nucleus nuclear potential model, respectively}
\label{NuclearModel}
\begin{tabular}{c|c|c|c|c|c|c|c}
\hline
 & $^{25}$Mg$^+$ & $^{43}$Ca$^+$ & $^{87}$Sr$^+$ & $^{137}$Ba$^+$ & $^{63}$Cu & $^{107}$Ag & $^{197}$Au  \\
\hline
RASSCF+DKH2 (PN)\textsuperscript{\emph{a}} & 579 &  852 & 1253 &  6286 &  6102 & 2300 &  6564 \\ \hline
RASSCF+DKH2 (FN)\textsuperscript{\emph{a}}& 579 &  850 & 1237 &  5955 &  6077 & 2245 &  5785 \\ \hline
RASSCF+X2C (FN)\textsuperscript{\emph{a}} & 554 &  749 &  899 &  3262 &  4703 & 1383 &  2488 \\ \hline
Lit. DFT(BP86)+DKH2 (PN)\textsuperscript{\emph{b}} & 601 & 813 & 1061 & 4570 & 6590 & 1959 & 3646 \\ \hline
Lit. DFT(BP86)+DKH2 (FN)\textsuperscript{\emph{c}} & - & - & - & - & 6556 & 1701 & 2922 \\ \hline
Lit. OOMP2+DKH2 (PN)\textsuperscript{\emph{d}} & 596 & 826 & 1042 & 4367 & 6933 & 2005 & 3586 \\ \hline
Lit. OOMP2+DKH2 (FN)\textsuperscript{\emph{d}} & 595 & 820 & 1015 & 4071 & 6834 & 1919 & 2910 \\ \hline
Lit. MCDF (FN)\textsuperscript{\emph{e}} &   -  &  -   &  -    &  -    & 5877 & 1724 & 3064 \\ \hline
Lit. 4CC (PN)         & 594\textsuperscript{\emph{f}} & 793\textsuperscript{\emph{g}} & 1000\textsuperscript{\emph{h}} & 4073\textsuperscript{\emph{i}}  & - & - & - \\ \hline
Experiment               &  596\textsuperscript{\emph{j}} & 806\textsuperscript{\emph{k}} &  990\textsuperscript{\emph{l}} & 4019\textsuperscript{\emph{m}} & 5867\textsuperscript{\emph{n}} & 1713\textsuperscript{\emph{o}} & 3054\textsuperscript{\emph{o}} \\ \hline
\end{tabular}
\\
\raggedright
\textsuperscript{\emph{a}} \footnotesize{This work: using DKH2 or X2C hyperfine Hamiltonian with RAS(11,5-1-24), RAS(19,9-1-24), RAS(37,18-1-18), RAS(45,22-1-9), RAS(29,14-1-18), RAS(47,23-1-9), and RAS(47,23-1-16), for Mg$^+$, Ca$^+$, Sr$^+$, Ba$^+$, Cu, Ag, and Au, respectively;}
\textsuperscript{\emph{b}} \footnotesize{Ref.~\citenum{Malkin2004}}
\textsuperscript{\emph{c}} \footnotesize{Ref.~\citenum{Malkin2006}}
\textsuperscript{\emph{d}} \footnotesize{Ref.~\citenum{Sandhoefer2013};}
\textsuperscript{\emph{e}} \footnotesize{Ref.~\citenum{Song2007};}
\textsuperscript{\emph{f}} \footnotesize{Ref.~\citenum{Sur2005};}
\textsuperscript{\emph{g}} \footnotesize{Ref.~\citenum{Sahoo2003};}
\textsuperscript{\emph{h}} \footnotesize{Ref.~\citenum{Martensson2002};}
\textsuperscript{\emph{i}} \footnotesize{Ref.~\citenum{Sahoo2003b};}
\textsuperscript{\emph{j}} \footnotesize{Ref.~\citenum{Itano1981};}
\textsuperscript{\emph{k}} \footnotesize{Ref.~\citenum{Arbes1994};}
\textsuperscript{\emph{l}} \footnotesize{Ref.~\citenum{Sunaoshi1993};}
\textsuperscript{\emph{m}} \footnotesize{Ref.~\citenum{Blatt1981};}
\textsuperscript{\emph{n}} \footnotesize{Ref.~\citenum{Ting1957};}
\textsuperscript{\emph{o}} \footnotesize{Ref.~\citenum{Wessel1953}.}
\end{table}

We analyze the effect of the nuclear electrostatic potential model on the HFC parameter. Table~\ref{NuclearModel} shows the magnitude of $A_\text{iso}$ calculated using both point-nucleus and finite-nucleus models for the nuclear electrostatic potential. In the finite-nucleus model, the nuclear charge distribution is represented by a single $s$-type Gaussian function. We show results obtained with the DKH2 relativistic hyperfine Hamiltonian that explicitly depends on the nuclear electrostatic potential (see Eqs.~\ref{Hdkh2-MHf}-\ref{wapp}). We emphasize that for both models of the nuclear electrostatic potential, the HFC operator is derived assuming a point-nucleus model for the magnetic vector potential. For each atom and ion we use the largest considered RASSCF active space. In particular, for Mg$^+$, Ca$^+$, Sr$^+$, Ba$^+$, Cu, Ag, and Au we use RAS(11,5-1-24), RAS(19,9-1-24), RAS(37,18-1-18), RAS(45,22-1-9), RAS(29,14-1-18), RAS(47,23-1-9), RAS(47,23-1-16), respectively.  For lighter ions like $^{25}$Mg$^+$ or $^{43}$Ca$^+$, the effect of the finite-nucleus model for the nuclear electrostatic potential is negligible. However, as the size of the nucleus increases for heavier atoms, this effect becomes more and more important. In particular, for $^{197}$Au the finite nucleus correction is as large as 13\%. In general, the $|A_\text{iso}|$ value calculated using a finite-nucleus electrostatic potential is lower compared to the point-nucleus value. This is in agreement with other theoretical works\cite{Malkin2006,Malkin2011,Sandhoefer2013} (see also Table~\ref{NuclearModel}). Such an effect can be understood by observing that for a finite-nucleus charge distribution an electron in the close proximity of the nucleus feels a smaller effective nuclear charge than it would feel for the point nucleus model. This leads to a weaker attraction and, thus, lowers electron spin density at the center of the nucleus, which reduces the Fermi contact contribution to the HFC parameter.\cite{Malkin2006,Malkin2011,Sandhoefer2013}

\begin{figure}[t!]
\centering
\includegraphics[width=1.0\linewidth]{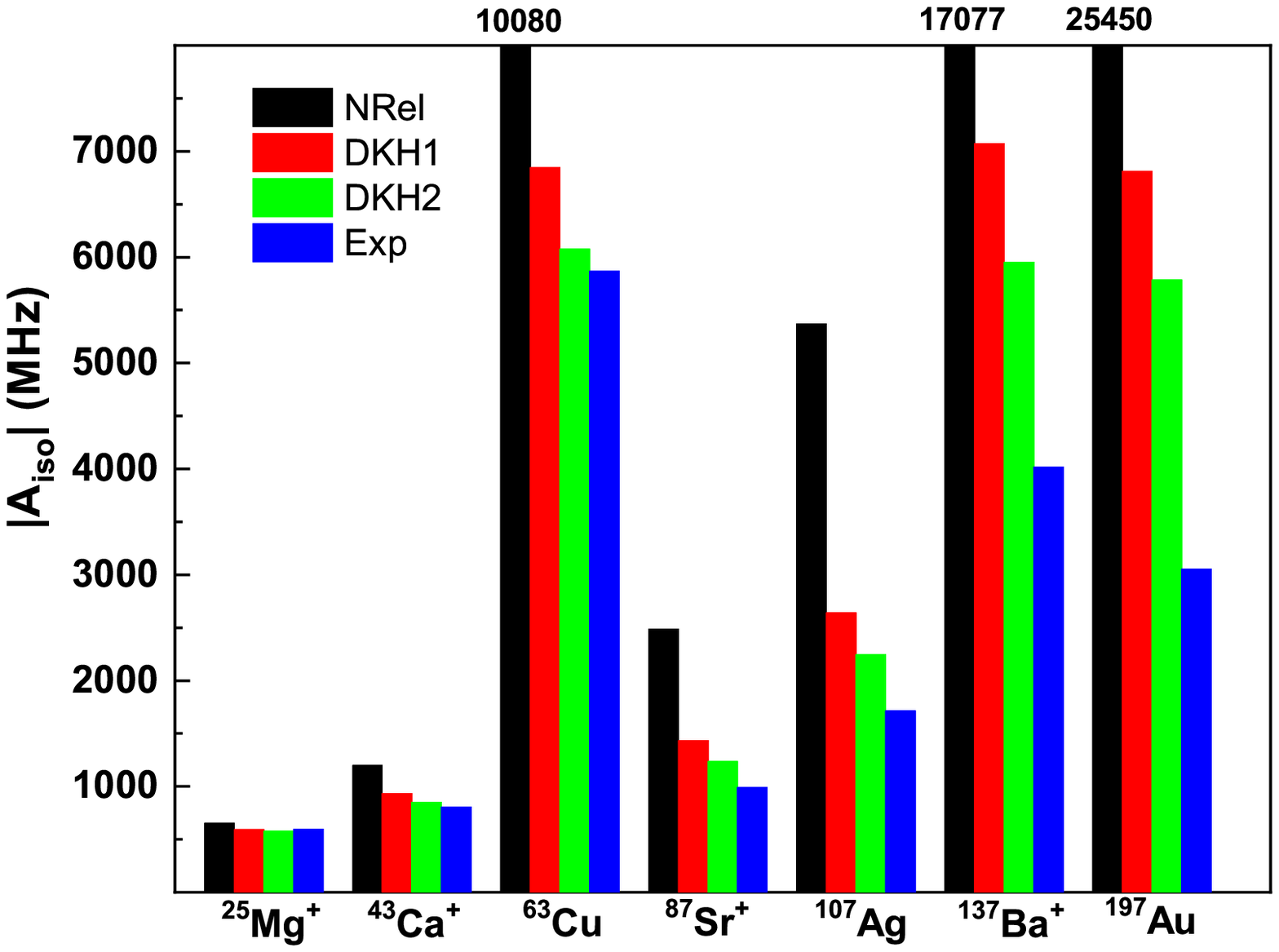}
\caption{Magnitude of the isotropic HFC constant for group-II cations and group-XI atoms calculated using the RASSCF method and the finite-nucleus model for the electrostatic nuclear potential. For Mg$^+$, Ca$^+$, Sr$^+$, Ba$^+$, Cu, Ag, and Au we use RAS(11,5-1-24), RAS(19,9-1-24), RAS(37,18-1-18), RAS(45,22-1-9), RAS(29,14-1-18), RAS(47,23-1-9), RAS(47,23-1-16), respectively. Black, red, and green colors denote nonrelativistic, DKH1, and DKH2 calculations, respectively. Experimental values\cite{Itano1981,Arbes1994,Sunaoshi1993,Blatt1981,Ting1957,Wessel1953} are in blue.}
\label{AtomsRes}
\end{figure}

We now discuss the importance of relativistic effects. Figure~\ref{AtomsRes} shows the magnitude of $A_\text{iso}$ calculated using different relativistic treatments of the hyperfine Hamiltonian (nonrelativistic, DKH1 and DKH2). For each atom and ion, we use the finite-nucleus electrostatic potential and the largest considered RASSCF active space (see the caption of Fig.~\ref{AtomsRes}). The calculations are compared to experimental values.\cite{Itano1981,Arbes1994,Sunaoshi1993,Blatt1981,Ting1957,Wessel1953}
Overall, the utilization of the relativistic hyperfine Hamiltonian reduces the $|A_\text{iso}|$ value, leading to a better agreement with experiment. 
This is due to the fact that the relativistic treatment of the hyperfine interaction introduces kinetic factors that regularize the singularity of the Dirac delta function in the nonrelativistic Fermi contact operator. For light ions such as Mg$^+$, the relativistic treatment has a rather small effect but nevertheless it improves the agreement with experiment. As the atomic number $Z$ increases, the relativistic effects become much more important. Even for $^{63}$Cu atom the use of the relativistic hyperfine Hamiltonian is essential. For atoms with larger $Z$, the difference between the DKH1 and DKH2 treatments also increases and the DKH2 correction further reduces the $|A_\text{iso}|$ value.

Deviations from the experimental data increase with the atomic number and the calculations typically overestimate the measured values\cite{Itano1981,Arbes1994,Sunaoshi1993,Blatt1981,Ting1957,Wessel1953} even in the DKH2 case. For most atoms and ions, however, the deviations 
are small and a good agreement with experimental data is achieved. Only in the case of the heaviest atoms/ions like $^{137}$Ba$^+$ and $^{197}$Au, significant deviations from the experimental values are found. In particular, for $^{197}$Au, the calculated DKH2 value overestimates the experimental value almost by a factor of two. One possible reason for this deviation is the point-nucleus approximation used for the nuclear magnetic vector potential. Indeed, the use of the finite-nucleus model for the nuclear magnetic moment distribution becomes more important for large $Z$ nuclei and has an overall effect of reducing $|A_\text{iso}|$ since the smeared nuclear magnetic moment locally probes a smaller spin density on average.\cite{Malkin2006} Another possible source of error is the neglect of the DKH transformation for two-electron repulsion integrals. This type of a picture-change effect\cite{Autschbach2012} is important for heavier nuclei and is expected to reduce $|A_\text{iso}|$, as pointed out by Ref.~\citenum{Malkin2004}. Furthermore, higher-order DKH corrections and additional dynamic correlations not included in our RASSCF calculations can contribute to the deviations from the experimental data.

We compare our best calculated DKH2 values to the X2C values calculated using modules implemented by Autschbach group\cite{Feng2021} in the OpenMolcas package\cite{Openmolcas} using the same active space (see Table~\ref{NuclearModel}). The calculated X2C values for the group XI neutral atoms are consistent with the results from Ref.~\citenum{Birnoschi2022}. Overall, the X2C values are lower compared to the DKH2 values and they underestimate the experimental data, although they seem to have better agreement with experimental values for heavier nuclei. Next, we compare our DKH2-calculated $|A_\text{iso}|$ values with the DFT+DKH2 (Refs.~\citenum{Malkin2004,Malkin2006}) and OOMP2+DKH2
(Ref.~\citenum{Sandhoefer2013}) results. For lighter ions such as $^{25}$Mg$^+$ and $^{43}$Ca$^+$, our results agree well with both DFT+DKH2 and OOMP2+DKH2 calculations, while for heavier ions like $^{87}$Sr$^+$, our results show somewhat larger deviations from the experimental values than the DFT+DKH2 and OOMP2+DKH2 values. On the other hand, for $^{63}$Cu, our DKH2 value agrees much better with experiment than both DFT+DKH2 and OOMP2+DKH2 results. However, for $^{107}$Ag, $^{137}$Ba$^+$, and $^{197}$Au  our results are much more deviated from the experimental data than the  DFT+DKH2 and OOMP2+DKH2 values. Lastly, for light and moderately heavy atomic systems, our  DKH2 values agree well with the values from the multiconfigurational Dirac-Fock theory (MCDF)\cite{Song2007} and the four-component coupled-cluster (4CC) method\cite{Sur2005,Sahoo2003,Martensson2002,Sahoo2003b} (see Table~\ref{NuclearModel}). However, deviations of our DKH2 values from those from latter two approaches increase as atomic number increases.

\subsection{\label{sec:diatom}Diatomic Molecules}

In order to study the relativistic effects on the dipolar contribution to the hyperfine interaction, we consider several simple diatomic radicals such as ScO, PdH, LaO, and HgF compounds. Considering cylindrical symmetry of the radicals, we can write the HFC matrix
as follows:
\begin{equation}
\mathbf{A}=
\begin{pmatrix}
A_\perp & 0       & 0 \\
0           & A_\perp & 0 \\
0           & 0       & A_\parallel
\end{pmatrix}.
\end{equation}
Here, $A_\parallel$ and $A_\perp$ denote the principal components of the HFC matrix in the directions parallel and perpendicular to the molecular axis, respectively. We focus on spin-$1/2$ radicals with orbitally nondegenerate ground states so that the PSO contribution is negligible and the Fermi contact contribution is described by the isotropic part of the HFC matrix $A_\text{iso}=\text{Tr}(\mathbf{A})/3$. The dipolar contribution, on the other hand, is described by the following quantity $A_\text{dip}=(A_\parallel-A_\perp)/3$.

Table~\ref{SimpleMolecules} shows the calculated $A_\text{iso}$ and $A_\text{dip}$ parameter values and the corresponding experimental values for the considered molecules. For calculations we use experimental or optimized bond lengths of 1.668~\AA,~1.529~\AA,~1.826~\AA,~and 2.077~\AA~for ScO\cite{Radzig1985}, PdH\cite{Radzig1985}, LaO\cite{Radzig1985}, and HgF\cite{Belanzoni2001}, respectively. A finite-nucleus model for the nuclear electrostatic potential is used. The calculations are performed using the RASSCF method with RAS(29,14-1-13), RAS(29,14-1-18), RAS(31,15-1-15), and RAS(35,17-1-17) for ScO, PdH, LaO and HgF, respectively. The detailed study of convergence of the HFC parameters as a function of active space size is described in Table~S7-S10 of the SI.

\begin{table}[t]
\centering
\caption{Calculated HFC parameters (MHz) for different diatomic radicals compared with experimental data and other calculations in the literature (Lit.). Here, $A_\text{iso}=\text{Tr}(\mathbf{A})/3$ and $A_\text{dip}=(A_\parallel-A_\perp)/3$} 
\label{SimpleMolecules}
\begin{tabular}{c|c|c|c|c|c|c|c|c|c|c}
\hline
Molec. & \multicolumn{5}{|c|}{$|A_\text{iso}|$} & \multicolumn{5}{|c}{$|A_\text{dip}|$} \\
\cline{2-11}
    Center  & NRel\textsuperscript{\emph{a}}  & DKH1\textsuperscript{\emph{a}} & DKH2\textsuperscript{\emph{a}} & Lit. & Exp & NRel\textsuperscript{\emph{a}} & DKH1\textsuperscript{\emph{a}} & DKH2\textsuperscript{\emph{a}} & Lit. & Exp \\
\hline
ScO  & 2652   & 2077  & 1917  & 1991\textsuperscript{\emph{b}}  & 1947\textsuperscript{\emph{d}}  
& 23 & 23  & 23  & 20\textsuperscript{\emph{b}}  & 25\textsuperscript{\emph{d}}  \\
$^{45}$Sc & & & & 1930\textsuperscript{\emph{c}} & & & & & 20\textsuperscript{\emph{c}} & \\ \hline
PdH  & 2462   & 1234  & 1050  & 910\textsuperscript{\emph{c}} & 823\textsuperscript{\emph{f}}   
& 29   & 27  & 27  & 29\textsuperscript{\emph{c}}  & 22\textsuperscript{\emph{f}}  \\
$^{105}$Pd & & & & 872\textsuperscript{\emph{e}} & & & & & 39\textsuperscript{\emph{e}} & \\ \hline
LaO  & 13729  & 5588  & 4704 & 3660\textsuperscript{\emph{c}}  & 3662\textsuperscript{\emph{d}}
&  16  & 13  & 13  & 19\textsuperscript{\emph{c}}  & 32\textsuperscript{\emph{d}} \\
$^{139}$La & & & & 2502\textsuperscript{\emph{g}} & & & & & 14\textsuperscript{\emph{g}} & \\  \hline
HgF & 236577 & 62092 & 52808 & 18195\textsuperscript{\emph{c}}  & 22127\textsuperscript{\emph{j}}
& 121  & 102 & 105 & 110\textsuperscript{\emph{c}}  & 247\textsuperscript{\emph{j}} \\
$^{199}$Hg & & & & 19469\textsuperscript{\emph{h}} & & & & & 86\textsuperscript{\emph{h}} & \\ 
 & & & & 20024\textsuperscript{\emph{i}} & & & & & - & \\
\hline
\end{tabular}
\\
\raggedright
\textsuperscript{\emph{a}}\footnotesize{This work: RAS(29,14-1-13), RAS(29,14-1-18), RAS(31,15-1-15), and RAS(35,17-1-17) are used for ScO, PdH, LaO and HgF, respectively;}
\textsuperscript{\emph{b}}\footnotesize{Ref.~\citenum{Pantazis2019}: B3LYP+NR;}
\textsuperscript{\emph{c}}\footnotesize{Ref.~\citenum{Belanzoni2001}: GGA+ZORA;}
\textsuperscript{\emph{d}}\footnotesize{Ref.~\citenum{Knight1999};}
\textsuperscript{\emph{e}}\footnotesize{Ref.~\citenum{Lan2015}: DMRG/CASSCF+DKH3;}
\textsuperscript{\emph{f}}\footnotesize{Ref.~\citenum{Knight1971};}
\textsuperscript{\emph{g}}\footnotesize{Ref.~\citenum{Knight1999}: B3LYP+NR;}
\textsuperscript{\emph{h}}\footnotesize{Ref.~\citenum{Malkin2006}: DFT(BP86)+DKH2;}
\textsuperscript{\emph{i}}\footnotesize{Ref.~\citenum{Feng2021}: RASSCF+X2C;}
\textsuperscript{\emph{j}}\footnotesize{Ref.~\citenum{Knight1981}}
\end{table}

The behavior of the Fermi contact contribution is similar to that of the atomic systems. Overall the Fermi contact contribution, i.e. the $|A_\text{iso}|$ value, is reduced by inclusion of the relativistic effects. Although most of the reduction is achieved at the level of the DKH1 correction, the DKH2 correction gives rise to a further decrease of $|A_\text{iso}|$ and it improves agreement with experiment. In fact, for ScO, the DKH2-included $|A_\text{iso}|$ value agrees well with experiment. On the other hand, for heavier nuclei in the radicals, the calculations tend to overestimate the experimental values even at the DKH2 level.

The value of $A_\text{dip}$ is much less sensitive to the relativistic treatment. Indeed, the dipolar contribution represents the through-space magnetostatic interaction of the nuclear spin with the electron spin density. The latter is primarily localized in the valence electron region where the scalar relativistic effects are less important. Consequently, the DKH treatment leads to only small corrections with respect to the nonrelativistic value, although the corrections increase for heavier nuclei. For example, for the $^{199}$Hg center in HgF, the DKH treatment changes the nonrelativistic $|A_\text{dip}|$ value by only 13\%. Similarly, the dipolar contribution is not significantly affected by the finite-nucleus treatment of the nuclear electrostatic potential.

For light or moderately heavy nuclei such as $^{45}$Sc or $^{105}$Pd in the considered radicals, the calculated $|A_\text{dip}|$ values agree well with the experimental data. However, the agreement with experiment deteriorates for heavier nuclei such as the $^{139}$La center in LaO or $^{199}$Hg center in HgF, and  the calculations underestimate the experimental $|A_\text{dip}|$ value almost by a factor of two. As discussed earlier, since the relativistic corrections are rather small, these errors are likely due to dynamic correlations which cannot be fully included within the RASSCF method for heavier atoms.

We compare our RASSCF+DKH2 results to the HFC parameters calculated by other groups using different \emph{ab initio} methods (see Table~\ref{SimpleMolecules}). For the $^{45}$Sc center in ScO, the GGA+ZORA calculations\cite{Belanzoni2001} of $|A_{\rm{iso}}|$ agree with our results, although our $|A_{\rm{dip}}|$ value shows a better agreement with the experiment.\cite{Knight1999} The nonrelativistic B3LYP calculations\cite{Pantazis2019} somewhat overestimate the $|A_{\rm{iso}}|$ value for the $^{45}$Sc center in ScO, even though a more accurate exchange-correlation functional is being used. This is another indication of an importance of relativistic treatment for the Fermi contact contribution. For the heavier $^{139}$Pd center in PdH, both our RASSCF+DKH2 and the GGA+ZORA calculations\cite{Belanzoni2001} overestimate the experimental\cite{Knight1971} value of $|A_{\rm{iso}}|$, although the GGA+ZORA result is closer to the experiment. Even better agreement with the experimental value of $|A_{\rm{iso}}|$ is achieved using the CASSCF/DMRG+DKH3 calculations with a large active space.\cite{Lan2015} On the other hand, the CASSCF/DMRG+DKH3 calculations significantly overestimate the value of $|A_{\rm{dip}}|$ for the $^{139}$Pd center in PdH compared to the experiment.\cite{Knight1971} Both our RASSCF+DKH2 value and the GGA+ZORA\cite{Belanzoni2001} result perform much better with our results being closer to the experimental value. For the heavy $^{139}$La center in LaO, our RASSCF+DKH2 calculations significantly overestimate the experimental value\cite{Knight1999} of $|A_{\rm{iso}}|$. The nonrelativistic B3LYP calculations from Ref.~\citenum{Knight1999} significantly underestimate the experimental value. On the other hand, the GGA+ZORA $|A_{\rm{iso}}|$ value\cite{Belanzoni2001} is in an excellent agreement with the experiment. For $|A_{\rm{dip}}|$, our RASSCF+DKH2 value, the B3LYP+NR, and GGA+ZORA results all significantly underestimate the experimental value\cite{Knight1999} with the latter being the closest to the experiment. For the $^{199}$Hg center in HgF, our RASSCF+DKH2 $|A_{\rm{iso}}|$ value overestimates the experiment\cite{Knight1981} almost a factor of two, while the GGA+ZORA,\cite{Belanzoni2001} DFT(BP86)+DKH2,\cite{Malkin2004} and RASSCF+X2C\cite{Feng2021} are all somewhat lower but significantly closer to the experimental value.\cite{Knight1981} Our RASSCF+DKH2 $|A_{\rm{dip}}|$ value agrees with GGA+ZORA calculations\cite{Belanzoni2001} while the DFT(BP86)+DKH2 $|A_{\rm{dip}}|$ value\cite{Malkin2004} is somewhat lower. All the calculations, however, significantly underestimate the experimental $|A_{\rm{dip}}|$.\cite{Knight1981}

\subsection{\label{sec:SMM}Single Molecule Magnets}

We now study more complex molecules. The hyperfine interaction of the lanthanide centers in lanthanide-based SMMs\cite{Thiele2014,Godfrin2017,Kundu2022,Ruskuc2022} has been of considerable interest for quantum information science applications. In this case, the CASSCF/RASSCF methods are especially well-suited to properly describe the multiconfigurational nature of the electronic structure. In addition, the orbitally degenerate ground state gives rise to a large orbital angular momentum and consequently a large PSO contribution to the hyperfine interaction.\cite{Wysocki2020} We consider large trivalent and divalent Tb-based SMMs such as Tb(III)Pc$_2$ (Ref.~\citenum{Branzoli2009}) and 
Tb(II)(Cp$^{\rm{iPr5}})_2$ (Ref.~\citenum{Gould2019}). The former and latter molecules consist of 113 and 111 atoms, respectively.

\begin{figure}[H]
\centering
\includegraphics[width=1.0\linewidth]{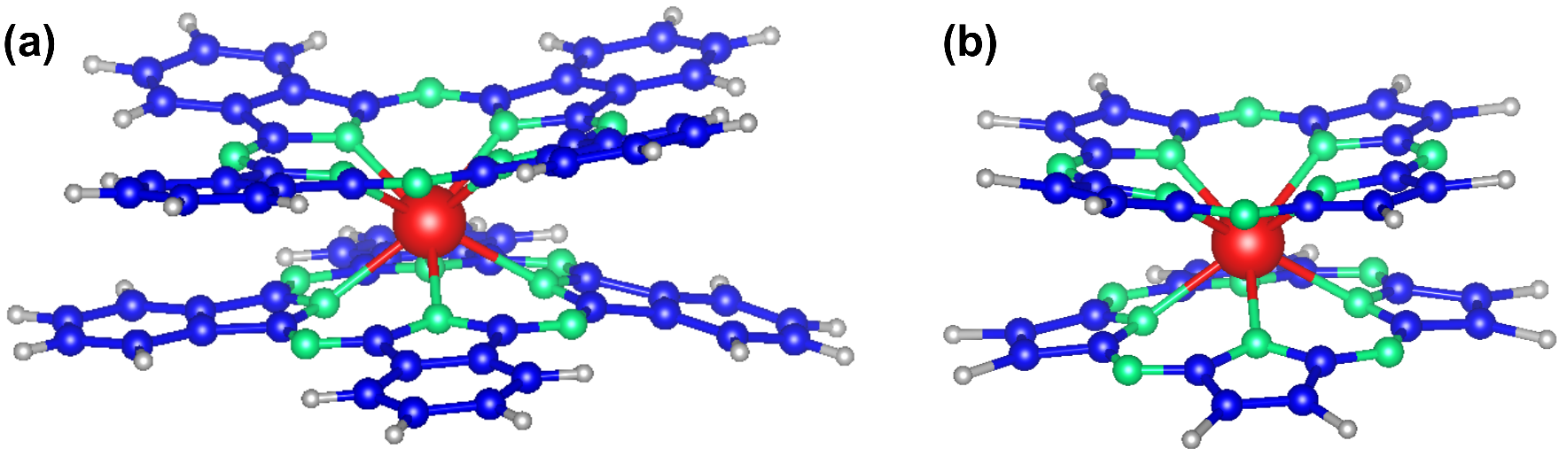}
\caption{(a),(b) Atomic structures of the anionic Tb(III)Pc$_2$ molecule without and with truncation of ligands. The structure without truncation (consisting of 113 atoms) is from experimental data, Ref.~\cite{Branzoli2009}. The truncation is the same as that in Ref.~\citenum{Feng2021}. Red, green, blue, and grey colors denote Tb, N, C, and H atoms, respectively. }
\label{SMMGeometries}
\end{figure}

The Tb(III)Pc$_2$ molecule is one of the first lanthanide-based molecules for which the SMM behavior was experimentally demonstrated.\cite{Ishikawa2003,Ishikawa2004,Ishikawa2004b} As shown in Fig.~\ref{SMMGeometries}(a), it comprises a trivalent Tb$^{3+}$ ion sandwiched between two approximately planar Pc ligands. For the anionic compound, unpaired electrons reside only in the Tb $4f$ orbitals. In the ground state, eight valence electrons in the $4f$ orbitals form the spin, orbital, and total angular momentum of $S=5$, $L=3$, and $J=6$, respectively, according to the Hund's rules. The $J=6$ ground multiplet is then split by the ligand fields. Among the split levels, the lowest-lying quasi-doublet is well separated from the higher-lying levels by 292 cm$^{-1}$.\cite{Pederson2019} Therefore, the ground quasi-doublet can be considered to be an effective spin-$\frac{1}{2}$ system.

We have studied the HFC parameter for such a system by using the nonrelativistic treatment of the hyperfine operators in Ref.~\citenum{Wysocki2020}. We showed that the hyperfine interaction of the system is strongly axial with the only $A_{zz}$ component being large (where the $z$ axis is along the magnetic easy axis), and that the dominant contribution originates from the PSO mechanism. The spin-dipole was found to be much smaller but significant with an opposite sign to the PSO contribution. For asymmetric TbPc$_2$ molecules (like the one from Ref.~\citenum{Branzoli2009}), there is non-negligible hybridization between Tb $f$ and $s$ orbitals which introduces a finite Fermi-contact contribution. These results are summarized in the column "NRel (full)" of Table~/ref{TbPc2-tab}, where a CASSCF(8,7) calculation with spin-orbit interaction included within the RASSI was used for the anionic TbPc$_2$ molecule without truncation. However, for more symmetric TbPc$_2$ molecules \cite{Ishikawa2003,Komijani2018}, the Fermi contact contribution is negligible because the spin density is primarily carried by the $4f$ electrons which have zero density at the Tb nucleus position.\cite{Wysocki2020}

\begin{table}[t]
\centering
\caption{Different contributions to the HFC parameter $A_{\rm{zz}}$ (MHz) for the anionic TbPc$_2$ molecule with and without truncation}
\label{TbPc2-tab}
\begin{tabular}{c|c|c|c|c|c|c|c}
\hline
      & NRel (full)\textsuperscript{\emph{a}} & DKH1 (full)\textsuperscript{\emph{b}}  & DKH2 (full)\textsuperscript{\emph{b}}  & NRel\textsuperscript{\emph{c}} & DKH1\textsuperscript{\emph{c}}  & DKH2\textsuperscript{\emph{c}}  & X2C\textsuperscript{\emph{d}}  \\ \hline
Total & 5994  & 6045 & 6030   & 6014 & 6065 & 6083 & 6117  \\ \hline  
FC+SD &  919  &  833 &  848   &  923 &  837 &  853 &  774   \\ \hline
FC    &  149  &   63 &   78   &  148 &   62 &   78 &    5 \\ \hline
SD    &  770  &    - &   -    &  775 & -    &  -   &  770  \\ \hline
PSO   & 6913  & 6878 &   -    & 6936 & 6902 &  -   & 6891  \\ \hline
\end{tabular}
\\
\raggedright
\textsuperscript{\emph{a}}\footnotesize{Ref.~\citenum{Wysocki2020}: CASSCF(8,7)+SO-RASSI with nonrelativistic hyperfine for full molecule without truncation;}
\textsuperscript{\emph{b}}\footnotesize{This work: CASSCF(8,7)+SO-RASSI with DKH1 or DKH2 for full molecule without truncation. The total DKH2 value is obtained by addition of the DKH2 FC+SD value to the DKH1 PSO value;}
\textsuperscript{\emph{b}}\footnotesize{This work: CASSCF(8,7)+SO-RASSI with nonrelativisitic, DKH1 or DKH2 for full molecule {\it with} ligand truncation same as Ref.~\citenum{Feng2021}. The total DKH2 value is obtained by addition of the DKH2 FC+SD value to the DKH1 PSO value;}
\textsuperscript{\emph{d}}\footnotesize{Ref.~\citenum{Feng2021}:CASSCF(8,7)+SO-RASSI with X2C hyperfine for truncated TbPc$_2$.}
\end{table}

Table~\ref{TbPc2-tab} lists calculated different contributions to the $A_{zz}$ element of the HFC matrix for the anionic TbPc$_2$ molecule with and without truncation, by using different relativistic treatment of the HFC operators and the point-nucleus model for the nuclear electrostatic potential. Figure~\ref{SMMGeometries} shows untruncated and truncated molecules.
The total relativistic treatment for the asymmetric molecule can be decomposed into two parts: (i) FC+SD contribution and (ii) PSO contribution. For asymmetric systems,
it is not straightforward to separate individual relativistic FC and SD contributions from
a relativistic FD+SD contribution. For simplicity, we separate a relativistic FC contribution from a relativistic FD+SD contribution by subtracting the corresponding nonrelativistic SD value from the latter, using the observation that the SD contribution is insensitive to the relativistic treatment.

With inclusion of the relativistic effects, the Fermi contact contribution is now reduced roughly by a factor of two, compared to the nonrelativistic case. Interestingly, the DKH1 correction greatly lowers the Fermi contact contribution, while the DKH2 correction slightly increases the contribution with respect to the former. For the PSO contribution, the DKH1 correction only slightly lowers the nonrelativistic value (about 0.5\%), and so we do not further compute the DKH2 contribution. 
This trend can be explained by the fact that the orbital moment density is localized in the valence electron region where the electrons are less influenced by the scalar relativistic effects. Overall, the relativistic effects introduce only small corrections to the HFC parameter for the TbPc$_2$ molecule. This result agrees well with the X2C calculation by Ref.~\citenum{Feng2021}. We expect 
that this is true, in general, for lanthanide systems where the spin density is carried by $4f$ electrons.

\begin{figure}[t!]
\centering
\includegraphics[width=1.0\linewidth]{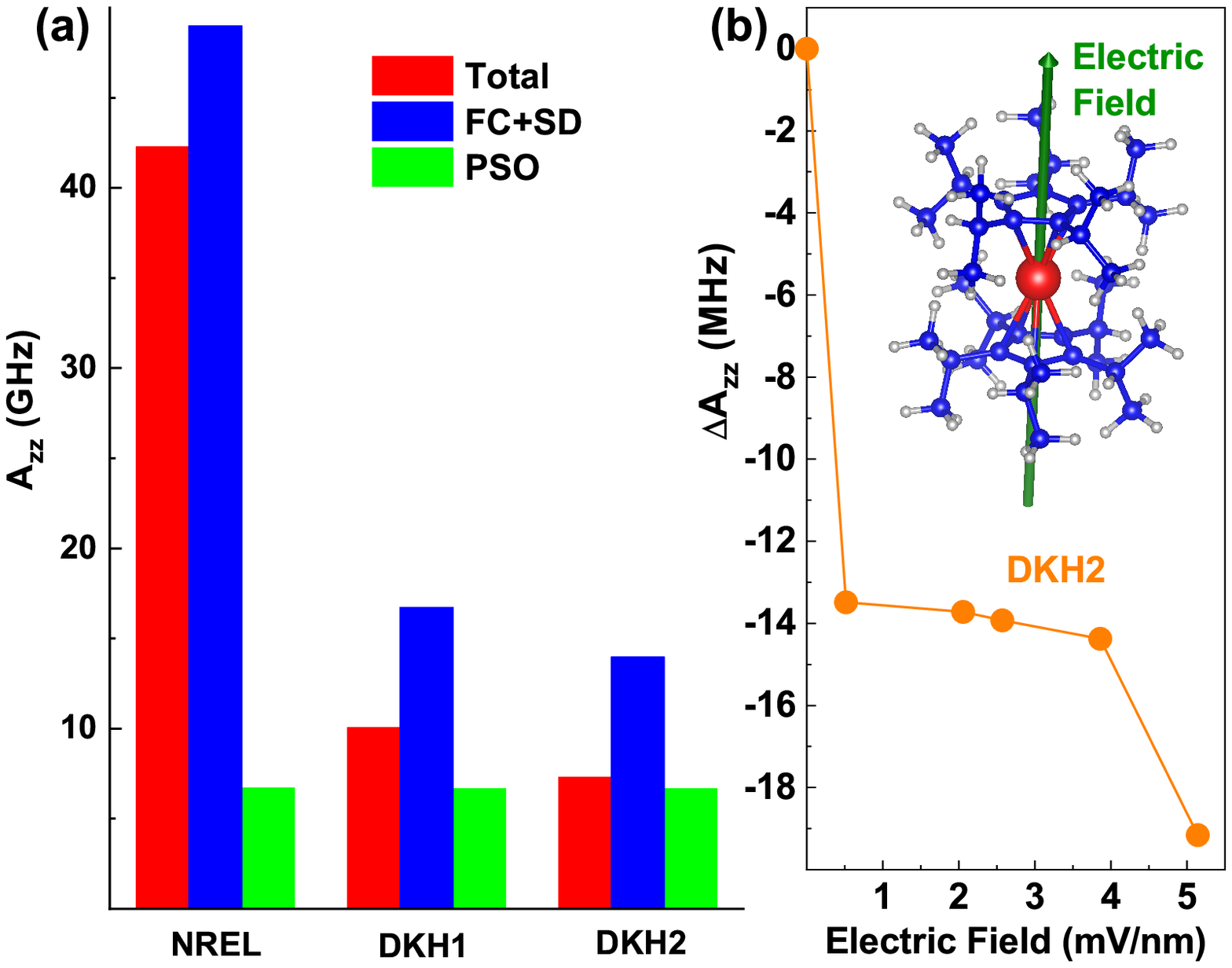}
\caption{HFC for the Tb(II)(Cp$^{\rm{iPr5}})_2$ molecule. (a) Total, FC+SD, and PSO contributions to the dominant $A_{zz}$ element for different relativistic treatments of the HFC Hamiltonian. (b) Electric-field dependence of the HFC parameter. Here, $\Delta A_{zz}$ is the difference between the $A_{zz}$ values obtained for finite and zero electric fields. The inset shows the molecular structure and the direction of an applied electric field.}
\label{TbCp2-fig}
\end{figure}

Let us now consider the Tb(II)(Cp$^{\rm{iPr5}})_2$ molecule.\cite{Gould2019} In contrast to the Tb(III)Pc$_2$ molecule, the Tb(II)(Cp$^{\rm{iPr5}})_2$ molecule consists of a divalent Tb ion (Tb$^{2+}$) with an extra Tb electron occupying a $6s/5d$ hybrid orbital, surrounded by two Cp$^{\rm{iPr5}}$ ligands (see the inset of Fig.~\ref{TbCp2-fig}(b)). We have investigated the electronic structure of this molecule using CASSCF(13,14)-SO-RASSI calculations, finding that this extra electron gives rise to a gigantic Fermi contact contribution to hyperfine interaction.\cite{Smith2020} Furthermore, we have predicted that the gigantic HFC parameter can be effectively tuned by an external electric field,\cite{Smith2020} which facilitates utilization of this molecule for quantum computing or sensing. However, the calculations in Ref.~\citenum{Smith2020} were performed using the nonrelativistic treatment of the HFC operators. As discussed above, such
a nonrelativistic formalism tends to significantly overestimate the Fermi contact contribution and, therefore, it would be interesting to investigate the effect of relativistic treatment on the HFC parameter for this molecule.

\begin{table}[t]
\centering
\caption{Different contributions to the HFC parameter (MHz) for Tb(Cp$^{\rm{iPr5}})_2$ calculated using different relativistic treatments of the HFC operators}
\label{TbCp2-tab}
\begin{tabular}{c|c|c|c|c}
\hline
      & NRel (finite-nucleus)\textsuperscript{\emph{a}} & NRel\textsuperscript{\emph{b}} & DKH1\textsuperscript{\emph{c}}  & DKH2\textsuperscript{\emph{c}}  \\ \hline
Total & 42322 & 42315 & 10071 &  7317  \\ \hline
FC+SD & 49020 & 49013 & 16734 & 13980 \\ \hline
FC    & 48220 & 48213 & 15933 & 13179   \\ \hline
SD    &   801 &   801 &  -    &  -    \\ \hline
PSO   &  6699 &  6700 &  6663 &  -     \\ \hline
\end{tabular}
\\
\raggedright
\textsuperscript{\emph{a}}\footnotesize{This work: CASSCF(13,14)-SO-RASSI for $S_{\rm{eff}}=1/2$ with the nonrelativistic HFC operators and finite-nucleus model;}
\textsuperscript{\emph{b}}\footnotesize{Ref.~\citenum{Smith2020}:CASSCF(13,14)-SO-RASSI for $S_{\rm{eff}}=1/2$ with the point-nucleus model;}
\textsuperscript{\emph{c}}\footnotesize{This work: CASSCF(13,14)-SO-RASSI for $S_{\rm{eff}}=1/2$ with the DKH1 or DKH2 correction and the point-nucleus model.}
\end{table}

Figure~\ref{TbCp2-fig}(a) and Table~\ref{TbCp2-tab} show different contributions to the HFC parameter for the Tb(Cp$^{\rm{iPr5}})_2$ molecule calculated using different relativistic treatments of the hyperfine operators. Here we use the point-nucleus model for the nuclear electrostatic potential, since the finite-nucleus effect turns out to be negligible for the nonrelativistic HFC operators.
The PSO contribution for this molecule does not significantly depend on the relativistic treatment 
of the HFC operators even at the DKH1 level. However, the Fermi contact contribution greatly reduces by consideration of the relativsitic correction at the DKH1 level, and it further decreases with the DKH2 correction. Importantly, even with inclusion of the relativistic correction, the Fermi contact contribution is very large, and it can be significantly varied by small external electric fields, as shown in Fig.~\ref{TbCp2-fig}(b). Therefore, while the HFC parameter is significantly smaller than the value predicted from the nonrelativistic calculations, the qualitative conclusion from Ref.~\citenum{Smith2020} remains valid.

\section{\label{sec:concl}Conclusions}

The relativistic treatment of the magnetic hyperfine Hamiltonian up to the DKH2 approximation has been implemented in the Molcas/OpenMolcas code. This implementation has been applied to atoms, diatomic molecules, and large Tb-based SMMs, in order to compute the relativistic HFC parameter by using CASSCF/RASSCF-SO-RASSI methods. For all the considered systems, when the relativitistic treatment is included, the Fermi contact contribution becomes greatly reduced compared to the nonrelativistic case. More specifically, the DKH1 correction is not negligible even for light atoms such as $^{25}$Mg$^+$, and it can reach an order of ten GHz for very heavy nuclei. For all the considered nuclei, the DKH2 correction is much smaller than the DKH1 correction, and it can be an order of GHz for very heavy nuclei. Overall, the DKH2-corrected Fermi contact contribution agrees well with experimental data for light and moderately heavy nuclei, whereas the deviation from experimental data increases as the atomic number increases. This deviation may be largely due to incomplete inclusion of dynamic correlation and neglect of the DKH tranformation for two-electron integrals. We showed that the spin-dipole contribution is not sensitive to the relativistic treatment. Regarding the large SMMs, the PSO contribution is dominant for the trivalent molecule, while the Fermi contact contribution is dominant for the divalent molecule. The HFC parameter
for the divalent molecule has been enormously decreased by the relativistic treatment due to the reduced Fermi contact term, while the HFC parameter for the trivalent molecule has not been changed much by the relativistic correction. We found that the divalent molecule would show the strong hyperfine Stark effect even when the relativistic HFC operators are considered, and we expect the similar behaviour for many divalent lanthanide-based systems.

\begin{acknowledgement}
This work was funded by the Department of Energy (DOE) Basic Energy Sciences (BES) grant No DE-SC0018326. Computational support by Virginia Tech ARC and San Diego Supercomputer Center (SDSC) under DMR060009N.
\end{acknowledgement}

\begin{suppinfo}
Details of the formalism for the second-order DKH method, active-space dependence of the Fermi contact contribution to the HFC parameter for the atomic systems, and active-space dependence of the Fermi contact and spin-dipole contributions to the HFC parameter for the diatomic systems. This material is available free of charge via the Internet at http://pubs.acs.org. 
\end{suppinfo}

\bibliography{refs}




\end{document}